\documentclass[a4paper,12pt]{article}
\usepackage{epsfig}
\date{\today}
 \normalsize

\newcommand{\bmat}{\left(\begin{array}}
\newcommand{\emat}{\end{array}\right)}
\newcommand{\be}{\begin{equation}}
\newcommand{\ee}{\end{equation}}
\newcommand{\bea}{\begin{eqnarray}}
\newcommand{\eea}{\end{eqnarray}}

\def\NPB#1#2#3{Nucl. Phys. B{#1} (19#2) #3}
\def\PLB#1#2#3{Phys. Lett. B{#1} (19#2) #3}

\def\PRD#1#2#3{Phys. Rev. D{#1} (19#2) #3}

\def\B{{\bf B}}
\def\inte{{\bf Z}}

\def    \be            {\begin{equation}}
\def    \ee            {\end{equation}}
\def    \bea           {\begin{eqnarray}}
\def    \eea           {\end{eqnarray}}
\def\eps{\epsilon}
\def\a{\alpha}

\def\d{\delta}

\def\k{\kappa}
\def\sig{\sigma}

\def\th{\theta}
\def\vt{\vartheta}

\begin{document}
\renewcommand{\thefootnote}{\fnsymbol{footnote}}
\rightline{IPPP/03/52} \rightline{DCPT/03/104}
\vspace{.3cm}
{\large
\begin{center}
{\bf Fermion Masses and Mixing in Intersecting Branes Scenarios}
\end{center}}
\vspace{.3cm}

\begin{center}
N. Chamoun$^{1,2}$, S. Khalil$^{3,4}$, and E. Lashin$^{1,4}$\\
\vspace{.3cm}
$^1$\emph{The Abdus Salam ICTP, P.O. Box 586, 34100 Trieste, Italy.}
\\
$^2$\emph{Physics Department, HIAST, P.O.Box 31983, Damascus, Syria.}
\\
$^3$ \emph{IPPP, Physics Department, Durham University, DH1 3LE, Durham, U.K.}
\\
$^4$\emph{Ain Shams University, Faculty of Science, Cairo 11566, Egypt.}
\end{center}
\vspace{.3cm}
\hrule \vskip 0.3cm
\begin{center}
\small{\bf Abstract}\\[3mm]
\end{center}
We study the structure of Yukawa couplings in intersecting
D6-branes wrapping a factorizable 6-torus compact space $T^6$.
Models with MSSM-like spectrum are analyzed and found to fail in
predicting the quark mass spectrum because of the way in which the
family structure for the left-handed, right-handed quarks and,
eventually, the Higgses is `factorized' among the different tori.
In order to circumvent this, we present a model with three
supersymmetric Higgs doublets which satisfies the anomaly
cancellation condition in a more natural way than the previous
models, where quarks were not treated universally regarding their
branes assignments, or some particular branes were singled out
being invariant under orientifold projection. In our model, the
family structures of all the standard model particles arise in one
of the tori and can naturally lead to universal strength Yukawa
couplings which accommodate the quark mass hierarchy and the
mixing angles.
\begin{minipage}[h]{14.0cm}
\end{minipage}
\vskip 0.3cm \hrule \vskip 0.5cm
\section{{\large \bf Introduction}}
Uncovering the nature and origin of the fermion families and the
observed pattern of fermion mass hierarchies and mixings is one of
the most fundamental issues in high energy physics.
In the framework of the standard model (SM), the vacuum
expectation value (VEV) of the Higgs field responsible for the
electroweak symmetry breaking generate the fermion masses through
Yukawa couplings. However, the SM does not address the origin of
these couplings, and the observed values for the fermion masses
are considered as initial `input' parameters \cite{abel3}. In addition, the
electroweak Cabbibo-Kobayashi-Maskawa (CKM) mixing matrix arising
from the matrices that diagonalize the up- and down-quarks mass
matrices is determined experimentally  to have, again, a
hierarchical structure \cite{abel3} where the third generation
mixing is mostly with the second generation rather than the first.
Something similar of hierarchies and mixings happens for the
neutrinos and huge efforts were done in order to understand this
`flavor problem' of the structure of the fermion masses and
mixing. Phenomenological studies considered``textures"
\cite{IbTo1} in the form of the mass matrices leading to
approximately correct relations, and attempts to understand the
presence of such``textures" then followed in different flavor
models \cite{IbTo6} or within Grand Unified Theories (see \cite{IbTo7}
and references therein).

Despite the insight which can be gained from these
phenomenological studies of the fermion mass matrices, arguably
the true resolution to the flavor problem lies in the domain of
the underlying fundamental theory of which the SM would be the low
energy effective theory. Since at present Superstrings/``M"--theory
is the only candidate for a truly fundamental quantum theory of
all interactions, studies of the flavor structure of the Yukawa
couplings within four-dimensional superstring models are well
motivated. In particular, the couplings of the effective
Lagrangian in superstring theory are in principle calculable and
not input parameters, which allows to address the flavor problem
quantitatively without introducing ad hoc assumptions. Indeed, the
structure of fermion masses has been studied in a number of
semi-realistic heterotic string models such as abelian $Z_n$
orbifolds \cite{abel7,abel8} which have a beautiful geometric
mechanism to generate a mass hierarchy \cite{abel9to10,abel11to12}
and the resulting renormalizable Yukawa couplings can be
explicitly computed \cite{abel13to14,abel15to16} as functions of
the geometrical moduli. An important result of such studies was to
demonstrate that the trilinear superpotential couplings at the
string scale are generally either zero or ${\cal O}(1)$, such that
they can provide a natural explanation for the top quark Yukawa
coupling \cite{kane6}, while other mechanisms utilizing
higher-dimensional non-renormalizable operators do generate the
lighter Yukawa couplings \cite{kane6,kane4and8}.

With the advent of Dirichlet D-branes in type II and type I string
theory, the phenomenological possibilities of string theory have
widened in several respects. Type I and Type IIB orientifold
models \cite{kane12,kane14,kane15to17,khalil33}, where the gauge
groups of the effective low energy Lagrangian arises from sets of
coincident D branes and where the matter fields arise from open
strings which must start and end upon D-branes, were proposed and
investigations into their general phenomenological features have
been possible. In \cite{khalil33,khalil35} a classification of the
matter fields has been extracted based on general grounds and
formulas for the soft terms and renormalizable Yukawa couplings
were derived. This has enabled a number of studies for the
patterns of soft breaking parameters
\cite{khalil33,kane19,kane20} and Yukawa textures
\cite{khalil,kane}. Nevertheless, the study of the structure of
the renormalizable Yukawa matrices and its viability within these
scenarios of D-branes at singularities has proved to be unable to
explain the experimental data, since they would generally lead to
a variant of the ``democratic" texture of Yukawa, and one has to
break this democracy by perturbative higher order effects or
non-renormalizable operators, the nature of which are still
unclear \cite{kane}. However, recent studies of the flavor
problem within `interscting D-branes' models
\cite{IbMo1and4,aldazabal1,IbGe,kokorelis,cevetic} seemed more
promising \cite{IbYu,IbTo}. In these models, chiral fields to be
identified with SM fermions live at different brane intersections
and there is a natural origin for the replication of quark-lepton
generations. In fact, most models are toroidal or orbifold
(orientifold) compactifications of Type II string theory with
Dp-brane wrapping intersecting cycles on the compact space, and
typically the branes would intersect a multiple number of times
giving rise to the family structure. Moreover, Yukawa couplings
between three chiral fields arise from open string instantons
stretching a worldsheet with triangle shape in whose vertices lie
the chiral fields. Each worldsheet contribute semiclassically to
the Yukawa coupling weighted by ${\rm e}^{-A}$, where $A$ is the
worldsheet area. This exponential weighting makes very natural
the appearance of hierarchies in Yukawa couplings of different
fermions with a pattern controlled by the size of the triangles.

Yet, the simple model presented in \cite{IbYu} which is based on
D$6$-branes wrapping cycles on an orientifold  of $T^2 \times T^2
\times T^2$ and has the chiral spectrum of the minimal
supersymmetric standard model (MSSM) does not really give
acceptable fermion masses. It leads to a mass spectrum of two
massless and one massive eigenvalue for the Yukawa matrices. This
reproduces the leading effect of one generation being much heavier
than the other two and, thus, should be considered only as a
starting point for a deeper phenomenological description where
small variations on the set up might give rise to smaller but
non-vanishing masses for the rest of quarks. In fact, we trace
this problem of mass degeneracy to the factorizable form $T^2
\times T^2 \times T^2$ of the compactified space and to the fact
that the family structures of the quark doublets, the quark
singlets and the Higgses arise, each, in one of the tori different
from the others. This
leads to a Yukawa matrix of special `factorizable' form ($a_ib_j$)
which has always two vanishing eigenvalues.

We argue in this paper that one can get more interesting Yukawa
structures assuming three generations of supersymmetric Higgses
$(H^u_i,H^d_i)_{i=1,2,3}$. This allows the generation of family
structures for the quark doublets, singlets and Higgses to take
place in only one of the tori. In fact, several models with three
families, including Higgses, have been constructed
\cite{abel20,abel} and were favored by the unification of the
gauge couplings in heterotic string.
Importantly for our analysis, having three families of Higgses
would allow easily to satisfy the RR tadpole cancellation
condition which requires the number of fundamentals to equal that
of antifundamentals for $SU(2)$. This is because the Higgses can
account for the six $SU(2)$ doublets needed to be added to the
three lepton left-handed doublets in order to equal the nine
antidoublets of the three families of chiral left-handed quark
color triplets $3(3,\bar{2})$. This offers a natural
solution to the anomaly cancellation condition without the need
to put the left-handed quarks in different brane intersections
\cite{IbGe,Alday}, or to assume some specific properties satisfied by
some of the branes \cite{IbYu}. Moreover, having three Higgs doublets
introduces more Yukawa couplings which introduces more flexibility in the
computation of the mass matrices \cite{abel}, hence one can accommodate
the observed quark masses and their mixing angles.

The structure of the paper is as follows. In the next section we
briefly review the different models for intersecting branes
leading to an MSSM-like spectrum and state what gauge symmetry
they have. Following this, we describe our model of three
supersymmetric Higgses and the way it satisfies the anomaly
cancellation condition. A brief discussion on how to determine the
string scale in this class of models is presented in section 3.
Section 4 is devoted for a detailed analysis for the quark masses and mixings.
Our conclusions are given in section 5.
\section{{\large Intersecting Branes Models}}
In this section we start with reviewing the construction of MSSM-like
models from intersecting D-branes. We also set some notation that we
will use through the paper.
The intersecting D-branes scenario offers an interesting way to get chiral
fermions. Consider a bunch of N D$p$-branes and another set of M
D$p$-branes ($p>3$) both containing the Minkowski space and
intersecting at some angle in the ($p-3$) extra dimensions. One
then gets massless chiral fermions transforming as $(N,\bar{M})$
under the gauge group $U(N) \times U(M)$ which allows to represent
the SM fermions. In addition, if the extra 6 dimensions are
compact, the intersection of a couple of branes is in general
multiple and the replication of generations is natural. Recently,
a particularly interesting class of models yielding `just' the
massless fermion spectrum of the SM was constructed
\cite{IbGe,IbYu}. These models consider D$6$-branes in type IIA string
theory compactified on a factorizable 6-torus $T^2 \times T^2
\times T^2$. One can wrap a D6-brane on a 1-cycle of each $T^2$ so
it expands a three-dimensional cycle on the whole $T^6$. We
denote the wrapping numbers of the D6$_a$-brane
on the i-th $T^2$ by $(n^i_a,m^i_a)$. If one minimizes the volume of these three-cycles
in their homology class, they are described by hyperplanes
quotiened by a torus lattice and this implies that the number of
times two branes $D6_a$ and $D6_b$ intersect in $T^6$ is given by
the signed intersection number
\be I_{ab}\ =\
(n_a^1m_b^1-m_a^1n_b^1)(n_a^2m_b^2-m_a^2n_b^2)(n_a^3m_b^3-m_a^3n_b^3)
\label{internumber} \ee

In addition to this, one performs an `orientifold' projection on the torus
represented by the product $\Omega \times R$, where $\Omega$ is
the worldsheet parity operator and $R$ is the reflection operator
with respect to one of the axis of the tori. The set of fixed
points under $\Omega \times R$ forms an orientifold plane, namely
a subspace of spacetime where the orientation of the string can
flip. This set has 8 components and corresponds to $O6$-planes
wrapped on the 3-cycle with wrapping numbers $(n_i,m_i)=(1,0)$.
Now each D-brane $\alpha$ has a mirror image under $\Omega \times R$
denoted by $\alpha^{\star}$. If the brane wraps a cycle
$[\Pi_\alpha]=(n^i_\alpha,m^i_\alpha)_{i=1,2,3}$ and
$\eps_\alpha^{(i)}$ represents the transversal distance of the
brane $\alpha$ from the origin in the i-th torus in clockwise
sense from the direction defined by $[\Pi_\alpha]$, then the
mirror image brane $\alpha^{\star}$ would wrap a cycle
$[\Pi_{\alpha^\star}]=(n^i_\alpha,-m^i_\alpha)_{i=1,2,3}$
 for rectangular tori, and the corresponding translation shift
 from the origin in the i-th torus is given by $\eps_{\alpha^\star}^{(i)}= -
 \eps_\alpha^{(i)}$.

There were, so far, two ways of embedding the standard model gauge
group into products of unitary and symplectic gauge groups, and
both ways used four stacks $a,b,c,d$ (and their orientifold
mirrors) of D$6$-branes called respectively the Baryonic, Left,
Right and Leptonic branes. Both methods (models) succeed in
getting an MSSM-like spectrum free of anomalies. However, in order
to do so, in the first model two left-handed quarks were doublets
and one left-handed quark was an antidoublet, while in the second
model one of the branes (b) was singled out being invariant under
the orientifold projection. We will summarize in the next subsection
the set up of these two models then we will present in the following
subsection our set up to generate the SM-like spectrum with the
aid of 3 supersymmetric Higgs doublets.
%
\subsection{{\large\bf Models with MSSM-like spectrum}}
In the first model (see \cite{IbGe} for details), and let's call
it model A, one gets initially the gauge symmetry \bea Model \ A:
\qquad U(3)\times U(2) \times U(1)\times U(1)\eea resulting from
the following number of branes in the corresponding stacks:
\[N_a=3,N_b=2,N_c=1,N_d=1\] and then one would embed $SU(3)_c$
into $U(3)$ and $SU(2)_L$ into $U(2)$. In order to yield the
desired SM spectrum, it is enough to select the wrapping numbers
$(n^i_\alpha,m^i_\alpha)$ for the four sets of D$6$-branes in such
a way that the intersection wrapping numbers are given by: \bea
I_{ab}\ & = &  \ 1 \ ;\ I_{ab*}\ =\ 2  \nonumber \\
I_{ac}\ & = &  \ -3 \ ;\ I_{ac*}\ =\ -3  \nonumber \\
I_{bd}\ & = &  \ 0  \ ;\ I_{bd*}\ =\ -3  \nonumber \\
I_{cd}\ & = &  \ -3 \ ;\ I_{cd*}\ =\ 3 \label{intersec2} \eea
all other intersections vanishing. The massless fermion spectrum
living at the intersections is shown in Table \ref{tableA}, where
the $N_R$ represents a right-handed neutrino and the hypercharge
generator is defined as:
\be Q_Y \ =\ {1\over 6} Q_a\ -\ {1\over 2} Q_c \ +\ {1\over 2} Q_d .
\label{hyperA} \ee
\begin{table}[htb] \footnotesize
\renewcommand{\arraystretch}{1.25}
\begin{center}
\begin{tabular}{|c|c|c|c|c|c|c|c|}
\hline Intersection &
 Matter fields  &   &  $Q_a$  & $Q_b $ & $Q_c $ & $Q_d$  & Y \\
\hline\hline (ab) & $Q_L$ &  $(3,\bar{2})$ & 1  & -1 & 0 & 0 & 1/6 \\
\hline (ab*) & $q_L$   &  $2( 3,2)$ &  1  & 1  & 0  & 0  & 1/6 \\
\hline (ac) & $U_R$   &  $3( {\bar 3},1)$ &  -1  & 0  & 1  & 0 & -2/3 \\
\hline (ac*) & $D_R$   &  $3( {\bar 3},1)$ &  -1  & 0  & -1  & 0 & 1/3 \\
\hline (bd*) & $ L$    &  $3(1,\bar{2})$ &  0   & -1   & 0  & -1 & -1/2  \\
\hline (cd) & $E_R$   &  $3(1,1)$ &  0  & 0  & -1  & 1  & 1   \\
\hline (cd*) & $N_R$   &  $3(1,1)$ &  0  & 0  & 1  & 1  & 0 \\
\hline \end{tabular}
\end{center} \caption{ Standard model spectrum and $U(1)$ charges
in the first model (A) \label{tableA} }
\end{table}

In this model one adopts the choice of splitting the left-handed quarks
into one quark represented by the intersection $(ab)$ and the other two
reprsented by $(ab^*)$ for consistency requirements. In fact, as
already mentioned, the RR tadpole cancellation condition, which is
stronger than the gauge anomaly cancellation condition, requires
the same number of doublets and antidoublets. This choice, then,
allows the left-handed quarks not to be universal under the $U(1)_b$
charge so that if two left quarks were $U(2)$ doublets and the
other one was $U(2)$ antidoublet, then taking the SM leptons as
$U(2)$ antidoublets allows to satisfy the requirement without the
need of extra doublets. As to the Higgs field sector, the Higgses
would come from the intersection between $b(b^*)$ and $c(c^*)$
branes, and there are four possible varieties of them
$(h_i,H_i)_{i=1,2}$ since we have two varieties of left quarks
$(Q_L,q_L)$ and two varieties of right quarks $(U_R,D_R)$.

The second model (see \cite{IbYu,IbTo} for details), to be called
model B, presents a slight variation where $N_b=1$ but $b^*$, the
mirror of $b$, lies on top of it $(b=b^*)$, so it can actually be
considered as a stack of two branes which, under $\Omega$
projection, yields a $USp(2) = SU(2)$ gauge group. So the initial
gauge group is \bea Model \ B: \qquad U(3)\times SU(2) \times
U(1)\times U(1)\eea With the following intersection numbers \be
\begin{array}{lcl}
I_{ab}\ = \ 3, \\
I_{ac}\ = \ -3, & & I_{ac^*}\ =\ -3, \\
I_{db}\ = \ 3, \\
I_{dc}\ = \ -3, & & I_{dc^*}\ =\ -3,
\end{array}
\label{intersecB}
\ee
and all the other intersection numbers
being zero, one gets the spectrum shown in Table \ref{mssmB}.
\begin{table}[h]
\begin{center}
\begin{tabular}{|c|c|c|c|c|c|c|}
\hline Intersection &
 SM Matter fields  & $SU(3) \times SU(2)$  &  $Q_a$   & $Q_c $ & $Q_d$  & Y \\
\hline\hline (ab) & $Q_L$ &  $3(3,2)$ & 1   & 0 & 0 & 1/6 \\
\hline (ac) & $U_R$   &  $3( {\bar 3},1)$ & -1  &  1  & 0 & -2/3 \\
\hline (ac*) & $D_R$   &  $3( {\bar 3},1)$ &  -1  & -1    & 0 &
1/3
\\
\hline (db) & $L$    &  $3(1,2)$ &  0    & 0  & 1 & -1/2  \\
\hline (dc) & $N_R$   &  $3(1,1)$ &  0  & 1  &  -1  &  0  \\
\hline (dc*) & $E_R$   &  $3(1,1)$ &  0  & -1  & -1    & 1 \\
\hline
\end{tabular}
\end{center}
 \caption{\small \label{mssmB} Standard model spectrum and $U(1)$
charges in the second model (B).}\end{table}
%
The $N_R$ denotes the right-handed neutrino and the hypercharge
generator is defined as $Q_Y = \frac 16 Q_a - \frac 12 Q_c - \frac
12 Q_d$. Notice that we do not have here the $Q_b$ anomaly condition
since doublets and antidoublets in $SU(2)$ are the same (there is no
$U(1)_b$ to differentiate between them). A particular class of
configurations satisfying the conditions (\ref{intersecB}) is
presented in Table \ref{wnumbersB1} and the intersections $(bc)$
and $(bc^*)$ can be identified with the MSSM Higgs particles
$H^u$, $H^d$.
\begin{table}[h]
\begin{center}
\begin{tabular}{|c||c|c|c|}
\hline
 $N_i$    &  $(n_\alpha^1,m_\alpha^1)$  &  $(n_\alpha^2,m_\alpha^2)$   & $(n_\alpha^3,m_\alpha^3)$ \\
\hline\hline $N_a=3$ & $(1,0)$  &  $(1 , 3)$ &
 $(1  ,  -3)$  \\
\hline $N_b=1$ &   $(0, 1)$    &  $ (1,0)$  &
$(0,-1)$   \\
\hline $N_c=1$ & $(0,1)$  &
 $(0,-1)$  & $(1,0)$  \\
\hline $N_d=1$ &   $(1,0)$    &  $(1,3 )$  &
$(1 , -3 )$   \\
\hline
\end{tabular}
\end{center}
 \caption{\small \label{wnumbersB1} D6-brane
wrapping numbers giving rise to the chiral spectrum of the MSSM in
the second model (B). }
\end{table}
\subsection{{\large\bf Models with 3 Supersymmetric Higgs doublets}}
As discussed above, the first model (A) treats the left quarks
differently as regards their location at the branes intersections.
Moreover, the intersection numbers (eq.\ref{intersec2}) are not
`symmetric' among the branes and their mirrors ({\it e.g.}, $I_{bd}=0$,
$I_{bd^*}=3$). As to the second model (B), although it also
reproduces an SM-like spectrum (with a right-handed neutrino),
it singles out one of the branes by requiring its
invariance under orientifold action ($b\equiv b^*$). The origin of
these assumptions lie, as already said, in the consistency
requirement that the number of fundamentals should be equal to the number
of antifundamentals even for $SU(2)$.

We are proposing now another way to satisfy this condition. We
consider, as in the first model (A), a stack of two branes $(b)$
giving rise to $U(2)$ gauge symmetry. We will treat the three
left-handed quarks universally and consider that we have chiral
quarks in $3(3,\bar{2})$ under $SU(3) \times SU(2)$. The full
model must contain then nine fields $(1,2)$, three of which
correspond to left-handed leptons. As to the remaining six
doublets, we do not need extra doublets to be accounted for if we
take the natural assumption of three generations of Higgses
$(H^u_i,H^d_i)_{i=1,2,3}$. In fact, in both models (A) and (B) the
u-Higgs and the d-Higgs are assigned opposite $U(1)$ charges so
that anomaly would not be affected by including them. However, no
reason prohibits them from having the same $U(1)_b$ charge, and
thus they can provide the extra six doublets necessary for anomaly
cancellation

Our model will be purely toroidal, with no orientifold projection
and so no mirror branes added. We shall consider that the SM
particles reside at the intersections amongst 4 stacks of branes
$N_a=3$, $N_b=2$, $N_c=1$, $N_d=1$. The intersection numbers
would be then \be
\begin{array}{lccclccc}
|I_{ab}|\ = \ 3 & \mathrm{representing} & Q_L &,& |I_{bc}|=3 & \mathrm{representing} & H^u &, \\
|I_{ac}|\ = \ 3 & \mathrm{representing} & U_R &,& |I_{bd}|=6=3+3 & \mathrm{representing} & H^d,L &, \\
|I_{ad}|\ = \ 3 & \mathrm{representing} & D_R &,& |I_{cd}|=3 &
\mathrm{representing} & E_R &,
\end{array}
\label{intersecC} \ee

In this third model, and let's call it model C, we used the fact
that the Higgs $H^d$ and the lepton $L$ have the same $SU(3)
\times SU(2) \times U(1)_Y$ quantum numbers, and so one can
consider getting both of them at the intersection of the same branes
(b and d). Requiring that the observed hypercharge generator is a
linear combination of the four $U(1)$'s one finds the following
general solution
\bea
\label{GenHypChar}
Q_Y &=& (\frac 23+\beta\alpha) Q_a + (\frac 12+\beta\alpha) Q_b + \alpha Q_c
+\gamma (1+\beta\alpha) Q_d \eea where $\beta^2=\gamma^2=1$
defined in Table \ref{tableC} and $\alpha$ is arbitrary. Notice
that with the choice $\alpha=-\frac{1}{2}$, $\beta=+1$,
$\gamma=1$, we get the hypercharge defined in equation
\ref{hyperA}
\begin{table}[htb] \footnotesize
\renewcommand{\arraystretch}{1.25}
\begin{center}
\begin{tabular}{|c|c|c|c|c|c|c|c|}
\hline Intersection &
 Matter fields  & $SU(3) \times SU(2)$  &  $Q_a$  & $Q_b $ & $Q_c $ & $Q_d$  & $Q_Y$ \\
\hline\hline (ab) & $Q_L$ &  $3(3,\bar{2})$ & 1  & -1 & 0 & 0 & 1/6 \\
\hline (ac) & $U_R$   &  $3( {\bar 3},1)$ &  -1  & 0  & $\beta$  & 0 & -2/3 \\
\hline (ad) & $D_R$   &  $3( {\bar 3},1)$ &  -1  & 0  & 0  & $\gamma$ & 1/3 \\
\hline (bd) & $ L$    &  $3(1,2)$ &  0   & +1   & 0  & $-\gamma$ & -1/2  \\
\hline (bd) & $ H^d_1$    &  $3(1,2)$ &  0   & +1   & 0  & $-\gamma$ & -1/2  \\
\hline (bc) & $ H^u_2$    &  $3(1,2)$ &  0   & +1   & $-\beta$  & 0 & 1/2  \\
\hline (cd) & $E_R$   &  $3(1,1)$ &  0  & 0  & $-\beta$  & $\gamma$  & 1   \\
\hline \end{tabular}
\end{center} \caption{ Standard model spectrum and $U(1)$ charges
in the third model (C) \label{tableC} }
\end{table}

We will give in section \ref{analysis} an example of D$6$-branes
wrapping numbers realizing the conditions (\ref{intersecC}). As can
be seen from Table \ref{tableC}, all the $U(1)$ gauge groups are
anomaly free. One should consider also mixed gauge and
gravitational anomalies. However, we expect that these anomalies
are cancelled by a generalized Green-Schwarz mechanism and that
three combinations of the $U(1)$'s would get massive with mass
roughly of the order of the string scale, while the hypercharge
$Y$ combination would stay massless. The symmetries whose gauge
bosons become massive would disappear as gauge symmetries from the
low-energy effective field theory, but remain as global symmetries
unbroken in perturbation theory. In this respect, $U(1)_a$ and
$U(1)_d$ represent respectively the global Baryonic and Leptonic
number symmetries.  However, assigning $Q_d$ charges to the Higgs $H^d$
might lead to a breaking of the Lepton number symmetry when the Higgs
acquires a VEV. Notice that in this model we do not have a
right-handed neutrino as a chiral fermion from intersecting
branes. Also, once we assume the Higgses $H^u,H^d$ come in a
number of generations equal to that of the SM particles then the
gauge anomalies would be cancelled automatically. Thus, in this model
there is no relation between the number of colors and the
number of families as it was the case in the previous models (A)
and (B) \cite{IbYu,IbGe}.
%
%
\section{{\large\bf String scale in the model with three Higgs doublets}}
The toroidal models are in general non-supersymmetric, and might
have tachyons at intersections. However, it is possible to vary
the compact radii in order to get rid of the tachyons. Also, one
can adjust the radii so that there is one massless scalar at any
given intersection, which means that one gets N=1 SUSY at that
specific intersection \cite{IbGe}. For instance in the MSSM-like
model (B), if the ratios of radii in the second and third torus
are equal then the {\it same} N=1 SUSY is preserved at all
intersections and the model is locally N=1 supersymmetric, having
an MSSM spectrum with a minimal Higgs set \cite{IbYu}.

For non-supersymmetric models, stabilization of the hierarchy of
the weak scale can be achieved by lowering the string scale down
to few TeV \cite{antonA,antonB}, while for supersymmetric models
there are several arguments in favor of string scales in
``intermediate" range $M_I \approx 10^{10-14}$ GeV \cite{khalil}.
 Such arguments provide an explanation to the observed experimental
neutrino masses \cite{khalil11}, or a means to attack the hierarchy
problem of unified theories in supergravity models by getting the gravitino
mass around the weak scale $m_{3/2}\approx M_W$ in a natural way
without invoking any hierarchically suppressed non-perturbative
effect \cite{khalil12}. Also, for intermediate string scale
scenarios, charge and color breaking constraints on the acceptable
region of parameter space for soft supersymmetry breaking terms
become less important \cite{khalil15,khalil16}. In addition, the
observed ultra high energy ($10^{20}eV$) cosmic rays can be
explained, for intermediate string scale, as products of long
lived massive string mode decays \cite{khalil17}.

In order to compute the string scale at which the running coupling
constants intersect in model (C) with three Higgs doublets, one
uses the one-loop running equation for the gauge coupling
\bea \frac{1}{\alpha_i(Q)}=\frac{1}{\alpha_i (M_Z)} +
\frac{b_i^{NS}}{2\pi}\ln\frac{M_{S}}{M_Z} +
\frac{b_i^{S}}{2\pi}\ln\frac{Q}{M_{S}} \ , \label{running} \eea
where $\alpha_i=g_i^2/4\pi$ with $i=2,3,Y$,  and $b_i$'s are the
coefficients of the $\beta$-functions, and where $M_Z$ represents
the overall non-supersymmetric scale while $M_S \approx 500 GeV$
represents an overall supersymmetric scale \cite{munoz}. On the
other hand, from eq.(\ref{GenHypChar}) we have the following
relation at the string scale $M_I$
\be
\frac{1}{\alpha_Y(M_I)} = \frac{\alpha^2}{\alpha^c_1(M_I)} +
\frac{(1+\beta\alpha)^2}{\alpha^d_1(M_I)} + \frac{(1/2
+\beta\alpha)^2}{\alpha_2(M_I)} + \frac{(2/3
+\alpha\beta)^2}{\alpha_3(M_I)}.
\label{hyperRel}
\ee
For the SM content with one
Higgs doublet, the non-supersymmetric $\beta$-functions are given
by $b_3^{NS}=7$, $b_2^{NS}=19/6$ and $b_Y^{NS}=-C^2\times 41/6$,
where $C$ is the normalization constant of the $U(1)_Y$
hypercharge ($C^2=3/5$ in $SU(5)$ GUT). As for the supersymmetric
$\beta$-functions, considering three supersymmetric generations of
standard particles, two Higgs doublets and an arbitrary number of
extra particles, we have
\bea b_3^S &=& 3-\frac{1}{2}n_3\ , \label{susybeta3}
\\
b_2^S &=& -1-\frac{1}{2}n_2\ , \label{susybeta2}
\\
b_Y^S &=& -C^2\times (11+q)\ , \label{susybeta1} \eea
where
\bea q=\sum_{i=1}^{n_1} Y_i^2+2\sum_{j=1}^{n_2}
Y_j^2+3\sum_{k=1}^{n_3} Y_k^2\ , \label{qus} \eea
and $n_1$, $n_2$, $n_3$ are the number of extra $SU(3)\times SU(2)$
singlets, $SU(2)$ doublets and $SU(3)$ triplets, respectively,
with masses close to $M_S$ and hypercharges $Y_l$. From eqs.(\ref{running})
and (\ref{hyperRel}), one finds
\begin{eqnarray}
\!\!\!\!\!\!&&\!\!\!\ln \frac{M_I}{M_S} =
\frac{1}{\left[(\frac{2}{3}+\alpha \beta)^2 b_3^{S} + (\frac{1}{2}
+\alpha \beta)^2 b_2^{S}-  b_Y^{S} \right]} \Big\{ 2\pi
\Big(\frac{1}{\alpha_Y(M_Z)} -\frac{\alpha^2}
{\alpha_1^c(M_I)}- \frac{(1+\beta \alpha)^2}{\alpha_1^d(M_I)}\nonumber\\
\!\!\!\!\!\!&& - \frac{(\frac{1}{2} +\alpha
\beta)^2}{\alpha_2(M_Z)} - \frac{(\frac{2}{3}+\alpha
\beta)^2}{\alpha_3(M_Z)}\Big) +\left(b_Y^{NS} -(\frac{1}{2}+\alpha
\beta)^2 b_2^{NS} - (\frac{2}{3}+\alpha \beta)^2 b_3^{NS}\right) \ln \frac{M_S}{M_Z}
\Big\}
\end{eqnarray}
We shall use the experimental values \cite{munoz39} $M_Z=91.187$
GeV, $\alpha_3(M_Z)=0.1184$, $\alpha_2(M_Z)=0.0338$,
$\alpha_Y(M_Z)=0.01016$ given in the ${\overline{MS}}$ scheme. The
fact that we have four extra Higgs doublets with respect to the
case of the MSSM means that  we should take $n_2=4$ and
$n_1=n_3=0$. Now, assuming that $\alpha=1/2$, $\beta=+1$ and
$C^2=3/5$, one obtains \be \ln \frac{M_I}{M_S}
=40.56 - \frac{0.17}{\alpha_1^c(M_I)} -
\frac{1.59}{\alpha_1^d(M_I)} - 1.89 \ln \frac{M_S}{M_Z}. \ee
Thus for $\alpha_1^c(M_I) \sim \alpha_1^d(M_I) \sim 0.1$, one
finds $M_I \simeq 10^{12}$ GeV. One could also check that the curves of $\alpha_2$ and $\alpha_3$
cross at approximately this intermediate scale which is, as emphasized above, an attractive possibility.
 However, we should mention here that in this class of toroidal or
 orientifold models there may exist extra chiral fields that would
 change the gauge couplings and might lead to a lower string
 scale. Although this possibility is indeed crucial in
 non-supersymmetric models in order to avoid any hierarchy between
 the string scale and the electroweak scale, it is not essential
 in our model with supersymmetric content. The model could still
 be consistent if these additional fields are decoupled from our
 spectrum or the possible threshold corrections are small \cite{antonB,anton97}.
%
\section{{\large \bf Analysis of Fermion Masses and Mixing}}
\label{analysis}
In \cite{IbYu}, it was shown that a Yukawa
coupling between fields at the intersections of factorizable
3-cycles $\Pi_a$, $\Pi_b$ and $\Pi_c$ on a factorizable $T^{6}$ is
given by \be Y_{ijk} = h_{qu} \sigma_{abc} \prod_{r = 1}^n \vt
\left[
\begin{array}{c}
\d^{(r)} \\ \phi^{(r)}
\end{array}
\right] (\k^{(r)}), \label{totalyuki2} \ee Here, each triplet of
intersection $(i,j,k)$ is described by the following multi-indices
\be
\begin{array}{cc}
i = (i^{(1)}, i^{(2)}, i^{(3)}) \in \Pi_a \cap \Pi_b ,
& \quad i^{(r)} = 0, \dots, |I_{ab}^{(r)}| -1, \\
j = (j^{(1)}, j^{(2)}, j^{(3)}) \in \Pi_c \cap \Pi_a,
& \quad j^{(r)} = 0, \dots, |I_{ca}^{(r)}| -1, \\
k = (k^{(1)}, k^{(2)}, k^{(3)}) \in \Pi_b \cap \Pi_c , & \quad
k^{(r)} = 0, \dots, |I_{bc}^{(r)}| -1,
\end{array}
\label{multiindices} \ee where ($r$) is an index indicating the
$r^{th}$ torus, and $I_{ab} = \prod_{r = 1}^n I_{ab}^{(r)} =
\prod_{r = 1}^n ( n_a^{(r)} m_b^{(r)} - n_b^{(r)} m_a^{(r)} )$,
where $I_{ab}^{(r)}$ denotes the intersection number of cycles $a$
and $b$ on the $r^{th}$ torus and $I_{ab}$ is the total
intersection number; $\sig_{abc} = {\rm sign }
(I_{ab}I_{bc}I_{ca})$, $h_{qu}$ stands for the quantum
contribution to the instanton amplitude and $\vt$ is the complex
theta function \be \vt \left[
\begin{array}{c}
\d \\ \phi
\end{array}
\right] (\k) = \sum_{l \in \inte} e^{\pi i (\d + l)^2 \k} \
e^{2\pi i (\d + l) \phi }. \label{thetacpx} \ee We have \bea
\d^{(r)} & = & \frac{i^{(r)}}{I_{ab}^{(r)}} +
\frac{j^{(r)}}{I_{ca}^{(r)}} + \frac{k^{(r)}}{I_{bc}^{(r)}} +
\frac{d^{(r)} \cdot \left(I_{ab}^{(r)} \eps_c^{(r)} + I_{ca}^{(r)}
\eps_{b}^{(r)} + I_{bc}^{(r)} \eps_{a}^{(r)}\right)} {I_{ab}^{(r)}
I_{bc}^{(r)} I_{ca}^{(r)}} + \frac{s^{(r)}}{d^{(r)}},
\label{paramT2ncpx1}\\
\phi^{(r)} & = & \left(I_{ab}^{(r)} \th_c^{(r)} + I_{ca}^{(r)}
\th_b^{(r)} + I_{bc}^{(r)} \th_a^{(r)}\right) / d^{(r)},
\label{paramT2ncpx2}\\
\k^{(r)} & = & \frac{J^{(r)}}{\a^\prime} \frac{|I_{ab}^{(r)}
I_{bc}^{(r)} I_{ca}^{(r)}|}{(d^{(r)})^2} \label{paramT2ncpx3} \eea
where $d^{(r)} = g.c.d. \left( I_{ab}^{(r)}, I_{bc}^{(r)},
I_{ca}^{(r)} \right)$ and $s^{(r)} \equiv
s(i^{(r)},j^{(r)},k^{(r)}) \in \inte$ is a linear function of the
integers $i^{(r)}$, $j^{(r)}$ and $k^{(r)}$. $\eps^{(r)}$
represents the `shifts' in the r$^{th}$ torus while the phase
$\theta^{(r)}$ accounts for adding Wilson lines around the D-brane
wrapping 1-cycles in the r$^{(th)}$ torus. $J^{(r)}$ represents
the Kahler structure of the r$^{(th)}$ torus and so $\k^{(r)}$
would be proportioanl to its area. Once, we have determined the
Yukawa couplings, one can compute the quark masses and
mixings to see whether the model reproduces the observed
hierarchical structure.
%
%
\subsection{{\large\bf Models with one supersymmetric Higgs doublet}}
In the first model (A) \cite{IbGe}, the case of a minimal set of
Higgs fields similar to the MSSM was shown to give masses to the
top, charm and bottom quarks while the strange, down, and up quarks
remained massless. It was argued that, with a double Higgs system case,
the observed hierarchy of fermion masses would be a consequence of the
different values of the Higgs fields and the hierarchical values
of Yukawa couplings, coming from geometrical considerations.

As to the seconde model (B) \cite{IbYu} and with the wrapping
numbers shown in Table \ref{wnumbersB1}, one gets one Higgs doublet
$H^u$($H^d$) at the intersection of $bc$($bc^*$). Yukawa
couplings of the form \be Y_{ij}^U Q_L^i H_u U_R^j, \quad \quad
Y_{ij*}^D Q_L^i H_d D_R^{j*}, \label{yukawas} \ee were computed,
and only the third generation of the quarks are massive. It was argued
that smaller perturbation of this set up can give rise to smaller
but non-vanishing masses for the rest of the quarks.

However, examining the model in depth would trace the problem of
having two zero eigenvalues in the Yukawa matrices to the
`factorizable' form that they take when the family replications for the
left-handed quarks and the right-handed quarks come from different
tori. For the case of Table \ref{wnumbersB1} we see that the index
$i^{(r)}_{ab} = 0, \dots, |I_{ab}^{(r)}| -1$, denoting the left
quarks, would span the values $0,1,2$ only in the 2$^{nd}$ torus,
while the index $i^{(r)}_{ac(c^*)} = 0, \dots, |I_{ac(c^*)}^{(r)}|
-1$, denoting the right quarks, would have its family structure only
in the 3$^{rd}$ torus. In such cases, and neglecting  Wilson
line effects, the Yukawa couplings would always be of the form \be
Y_{ij} \sim \vt^{(1)} \left[
\begin{array}{c}
\d(0) \\ 0
\end{array}
\right] \left( \k^{(1)} \right)
\times \vt^{(2)} \left[
\begin{array}{c}
\d(i) \\ 0
\end{array}
\right] \left( \k^{(2)} \right)
\times \vt^{(3)} \left[
\begin{array}{c}
\d(j) \\ 0
\end{array}
\right] \left( \k^{(3)} \right)
\ee
and so it is of a `factorizable' form \[ Y_{ij} \sim a_ib_j \]
Such matrices always
have two zero eigenvalues since, for instance, the second and third columns
are proportioanl to the first one.

Could we get more interesting phenomenology if the family
structures of both left-handed and right-handed quarks arise in
the same torus? The answer is no, if we restrict ourselves to one
Higgs doublet. In fact, if we adopt the wrapping numbers shown in
Table \ref{wnumbersB2}, where the branes $b$, $c$, $c^*$ are on
top of each other in the second torus, we find that the conditions
(\ref{intersecB}) are satisfied, and we have one massless
non-chiral Higgs doublet arising at the intersection of the brane
$b$ and the brane $c$ (or $c^*$) in the first and third tori. In
other words, there is a minimal Higgs sector with a
$\mu$-parameter determined by the distance between the branes $b$
and $c$ along the second torus.
\begin{table}[!hbp]
\begin{center}
\begin{tabular}{|c||c|c|c|}
 \hline
 $N_i$    &  $(n_\alpha^1,m_\alpha^1)$  &  $(n_\alpha^2,m_\alpha^2)$   & $(n_\alpha^3,m_\alpha^3)$ \\
\hline\hline $N_a=3$ & $(1,0)$  &  $(1 , 3)$ &
 $(1  ,  0)$  \\
\hline $N_b=1$ &   $(0, 1)$    &  $ (1,0)$  &
$(0,-1)$   \\
\hline $N_c=1$ & $(1,1)$  &
 $(1,0)$  & $(1,1)$  \\
\hline $N_d=1$ &   $(1,0)$    &  $(1,3 )$  &
$(1 , 0 )$   \\
\hline \end{tabular}
\end{center}
 \caption{\small Alternative example of D6-brane
wrapping numbers in the second model (B) leading to a chiral
spectrum of the MSSM. The family structure of both the left-handed
and right-handed quarks arises in the second torus.
\label{wnumbersB2}}\end{table}

In order to compute the Yukawa structure, we now note that the
family structure for both the left-handed and the right-handed
quarks is originated in the second torus, and so, neglecting Wilson
lines effect, we shall get \be Y_{ij} \sim \vt^{(1)} \left[
\begin{array}{c}
\d(0) \\ 0
\end{array}
\right] \left( \k^{(1)} \right) \times \vt^{(2)} \left[
\begin{array}{c}
\d^{(2)}(i,j) \\ 0
\end{array}
\right] \left( \k^{(2)} \right) \times \vt^{(3)} \left[
\begin{array}{c}
\d(0) \\ 0
\end{array}
\right] \left( \k^{(3)} \right)\ee where $\d^{(2)}(i,j) =
\frac{i+j}{3}+\lambda$, and $\lambda$ is a constant determined by
the shifts $\eps_a$, $\eps_b$, $\eps_c$. However, using the
periodicity of theta function \be \vt^{(r)} \left[
\begin{array}{c}
\d+1 \\ \phi
\end{array}
\right] \left( \k \right) = \vt^{(r)} \left[
\begin{array}{c}
\d \\ \phi
\end{array}
\right] \left( \k \right) \ee we get the following form for the
Yukawa matrix \be Y_{ij}  \sim \left(
\begin{array}{ccc}
a & b & c \\
b & c & a \\
c & a & b
\end{array}
\right)  \ee A matrix of this form has a spectrum such that two of
the eigenvalues are always opposite in sign, so it leads to two
degenerate states and can not reproduce the hierarchy in the
masses of the quarks. Having spotted the origin of the problem,
we now move to our third model (C).
%
\subsection{{\large \bf Model with 3 supersymmetric Higgs doublets}}
We saw in the previous subsection that having the family
structures of the left-handed quarks and the right-handed quarks
to arise from different tori leads to a mass matrix with two
vanishing eigenvalues. Also, having one Higgs doublet in the set
up would lead to a phenomenologically unacceptable form for the
mass matrices. One could also check that getting Higgs doublets
replication in one torus different
from the torus where the family structure arises for the quarks,
is similar to the one Higgs doublet situation. In this case, their effects
are factored out. Thus, we are led naturally to seek a situation
where we have more than one Higgs doublet and where the family
structures of all the left-handed quarks, right-handed quarks and
the Higgses arise in one of the tori. Recalling that the
assumption of three supersymmetric Higgses is also a `normal' choice in
order to cancel anomalies, we consider the model (C) with 3 Higgs
doublets. Since we are interested in this
paper only in the quark sector, we adopt the intersection numbers
(\ref{intersecC}) of the model (C) seeking to generate family
structures for the left and right quarks, as
well as for the Higgses, in the second torus, say. No
constraint is imposed on where the family structure for the leptons would
arise. This means that we are not trying here to interpret the
lepton masses and mixing, in particular that the model does not
contain right-handed neutrinos, nor Majorana neutrinos because
Lepton number is a symmetry, and so the question of neutrino
masses should be addressed differently.
\begin{table}[!hbp]
\begin{center}
\begin{tabular}{|c||c|c|c|}
\hline
 $N_i$    &  $(n_\alpha^1,m_\alpha^1)$  &  $(n_\alpha^2,m_\alpha^2)$   & $(n_\alpha^3,m_\alpha^3)$ \\
\hline\hline $N_a=3$ & $(1,0)$  &  $(2 , 3)$ &
 $(1  ,  0)$  \\
\hline $N_b=2$ &   $(0, 1)$    &  $ (1,3)$  &
$(0,1)$   \\
\hline $N_c=1$ & $(0,1)$  &
 $(1,0)$  & $(0,1)$  \\
\hline $N_d=1$ &   $(1,-1)$    &  $(3,3 )$  &
$(1 , -1 )$   \\
\hline \end{tabular}
\end{center}
\caption{\small Example of D6-brane wrapping numbers in the third
model (C). The family structure of the standard model particles
arises in the second torus. \label{wnumbersC}}
\end{table}

The conditions (\ref{intersecC}) are obtained with the wrapping
numbers shown in Table \ref{wnumbersC} where the branes $b$ and
$c$ are on top of each other in the first and third tori, and so
one gets massless non-chiral $H^u$ Higgs doublets. Moreover, we
see that the family structures of all the standard model
particles (the left and right handed quarks, the left and right
handed leptons and the Higgses) arise in the second torus. The
branes $b$ and $d$ give an intersection number equal to $6$, so
we can identify the first three intersections as the three $H^d$
Higgs doublets while the last three intersections would be the
three left lepton doublets. In this way the Higgs doublets $H^d$,
like the left leptons, are chiral but this does not lead to a
problem in constructing the MSSM superpotential since the chiral
$H^d$ can still form a $\mu$-term with the non-chiral $H^u$. An
approach to deal with chirality issues is to compactify over
$T^2/Z^2$ instead of $T^2$ \cite{aldazabal1}. However, one should
compute the detailed spectrum to check when this would
generically project onto chiral matter. We do not follow this
approach here, but instead, seek an assignment of wrapping
numbers that leads to chiral fermions. We found that such an
assignment could be obtained provided we allow multiwrapping
cycles for the branes. As an example, in Table \ref{wnumbersC} we
use a cycle $(3,3)$ for the brane $d$ in the second torus. Since
the wrapping numbers are not coprime the brane $d$ is
multiwrapped $3$ times over the cycle $(1,1)$ in this torus.
Normally, multiwrapping leads to an enhancement of the gauge
symmetry \cite{blumenhagen} (look also at
\cite{aldazabal1,aldazabal2} where similar multiwrapped
assignments were used for D$4$-branes, and hence for the whole
compact dimension of the brane). Nonetheless, in our case the
multiwrapping occurs only in the second torus. Even if this
partial multiwrapping in the second torus enhanced the
world-volume gauge group from $U(1)_d$ to $U(1)_d^3$ (with
generators $Q_d^a$, $a=1,2,3$), our discussion regarding the
hypercharge and the anomaly cancellation (equation
\ref{GenHypChar}) would stay valid with $Q_d=\sum_{a=1}^3 Q^a_d$
\cite{aldazabal2}. Thus, we shall not examine further the effects
of multiwrapping, especially that our model should be considered
as a step towards building a more realistic one. Actually, the
wrapping numbers in Table \ref{wnumbersC}, as it is the case in
Table \ref{wnumbersB1} and Table \ref{wnumbersB2} (see
\cite{IbYu,IbYu27} for discussion on this point), show that the
corresponding brane content by itself does not satisfy the RR
tadpole conditions $ \sum_a N_a \Pi_a = 0$, which would read as
follows
\begin{eqnarray}
\begin{array}{lcl}
\sum_a N_a n_a^1 n_a^2 n_a^3 = 0 & \quad &
\sum_a N_a n_a^1 m_a^2 m_a^3 = 0 \\
\sum_a N_a m_a^1 n_a^2 n_a^3 = 0 & \quad &
\sum_a N_a m_a^1 n_a^2 m_a^3 = 0 \\
\sum_a N_a n_a^1 m_a^2 n_a^3 = 0 & \quad &
\sum_a N_a m_a^1 m_a^2 n_a^3 = 0 \\
\sum_a N_a n_a^1 n_a^2 m_a^3 = 0 & \quad & \sum_a N_a m_a^1 m_a^2
m_a^3 = 0
\end{array}
\label{tadpole}
\end{eqnarray}
Yet, since tadpole cancellation conditions are closely connected
to cancellation of anomalies, and our model does cancel the
anomalies related to the gauge groups in Table \ref{tableC}, it is
not so surprising that with a slight change in the set up one
could satisfy the RR tadpole conditions. In fact, and in the
spirit of bottom to top approach, the model should be seen as a
submodel embedded in a bigger one where extra RR sources are
included. These may either involve some hidden, and possibly
non-factorizable, extra branes with no neat intersections with the
SM branes or some NS-NS background fluxes, with none of these
possibilities adding a `net' chiral matter content \cite{IbYu12}.
We shall not dwell on the details of such embedding which might
lead to extra matter, expectedly, heavy and disconnected from the
SM sector. Rather, we shall take our set up and examine what
interesting geometrical explanations for the fermion masses it
might lead to. The quark Yukawa coupling with the $H^d$ Higgs
would then be proportional to a product of three theta functions
(neglecting again the Wilson line phase) \be \vt^{(1)} \left[
\begin{array}{c}
\d(0) \\ 0
\end{array}
\right] \left( \k^{(1)} \right) \times \vt^{(2)} \left[
\begin{array}{c}
\d(i,j,k) \\ 0
\end{array}
\right] \left( \k^{(2)} \right) \times \vt^{(3)} \left[
\begin{array}{c}
\d(0) \\ 0
\end{array}
\right] \left( \k^{(3)} \right)\ee with $i,j,k=0,1,2$. The index
$k$ runs only over the first three $I_{bd}$ intersections
identified with the $H^d$ Higgs doublets. Thus the quark Yukawa
couplings for both the $H^u$ and $H^d$ Higgses would be
proportional to \be Y_{ijk} \sim \vt^{(2)} \left[
\begin{array}{c}
\d(i,j,k) \\ 0
\end{array}
\right] \left( \k^{(2)} \right) \ee and so we will restrict,
henceforth, our discussion to the second torus. For the U-quark
Yukawa coupling $Y^uH^uQ_LU_R$ we have
$|I_{ab}|=|I_{bc}^{(2)}|=|I_{ac}|=3$. This is similar to the case
of elliptic fibration discussed in \cite{IbYu,IbYu17} where the
intersection numbers are not coprime and only the triplets of
intersection satisfying the selection rule \be i + j + k \equiv 0
\ {\rm mod} \ 3. \label{selection} \ee are connected by an
instanton. We then get the following Yukawa couplings: \be
Y_{ij1} \sim \left(
\begin{array}{ccc}
A & 0 & 0 \\
0 & 0 & B \\
0 & C & 0
\end{array}
\right), \quad Y_{ij2} \sim \left(
\begin{array}{ccc}
0 & 0 & C \\
0 & A & 0 \\
B & 0 & 0
\end{array}
\right), \quad Y_{ij3} \sim \left(
\begin{array}{ccc}
0 & B & 0 \\
C & 0 & 0 \\
0 & 0 & A
\end{array}
\right), \label{yukilocal} \ee with \be A = \vt \left[
\begin{array}{c}
\eps/3 \\ 0
\end{array}
\right] (3 J/ \a'), \quad B = \vt \left[
\begin{array}{c}
(\eps - 1)/3 \\ 0
\end{array}
\right] (3 J/ \a'), \quad C = \vt \left[
\begin{array}{c}
(\eps + 1)/3 \\ 0
\end{array}
\right] (3 J/ \a'), \label{thetas} \ee and where we have $\eps =
\eps_a + \eps_b + \eps_c$. For the D-quark Yukawa coupling
$Y^dH^dQ_LD_R$ one would get the same result with a different
$\eps$-shift $\eps' = \eps_a + \eps_b + \eps_d$. However, as we
shall see, a numerically good fit is obtained around $\eps \simeq
\eps' \simeq 0$, and to fix the ideas let's take $\eps =\eps'$.
Thus, we get the U-quark mass matrix \be M^u_{ij}  = h_{qu} \left(
\begin{array}{ccc}
Av^u_1 & Bv^u_3 & Cv^u_2 \\
Cv^u_3 & Av^u_2 & Bv^u_1 \\
Bv^u_2 & Cv^u_1 & Av^u_3
\end{array}
\right)  \ee and the D-quark mass matrix \be M^d_{ij}  = h_{qu}
\left(
\begin{array}{ccc}
Av^d_1 & Bv^d_3 & Cv^d_2 \\
Cv^d_3 & Av^d_2 & Bv^d_1 \\
Bv^d_2 & Cv^d_1 & Av^d_3
\end{array}
\right)  \ee where $h_{qu}$ includes the quantum fluctuation
factor and we expect it to be similar for the u- and d- quarks
since leading effects would come from QCD loops \cite{IbTo}, and
$v^{u,d}_i$ is the VEV for the Higgs $H^{u,d}_i$ with \be
\sum_{i=1}^3(v^u_i)^2+(v^d_i)^2 = (174)^2 (GeV)^2
\label{higgscondition} \ee The quark masses are obtained by
diagonalizing the above mass matrices \be
\begin{array}{rcl}
U_LM^uU_R^{\dagger} & = &
d_U \\
D_LM^dD_R^{\dagger} &=& d_D
\end{array}
\label{matdef} \ee where $U_L$, $U_R$, $D_L$, $D_R$ are unitary
matrices and \[d_U=diag(m_t,m_c,m_u)\]\[d_D=diag(m_b,m_s,m_d)\]
while the CKM matrix is given by \be CKM = U_LD_L^{\dagger}
\label{ckmdef} \ee We have 7 free parameters consisting of the 6
Higgs VEVs with the constraint (\ref{higgscondition}), the area of
the torus and the shift $\eps$. We will not include the unknown
overall multiplicative factor $h_{qu}$ which is of order ${\cal
O}(1)$. This set of parameters can be fixed by the quark masses and
one mixing angle and the model has to predict the remaining two mixing
angles in the CKM. This might be a non-trivial task since one has to
span the whole range of all of these free parameters very carefully.
Here we will consider some examples and try to show that for a particular choice
of  these free parameters one may obtain a well studied Yukawa texture, like
for instance the Universal Strength Yukawa (USY) couplings (see Ref.\cite{branco}, and references
therein). Let us start with the case of approximately
symmetric matrices $M^u$, $M^d$ {\it i.e},  $B\approx C$. In this case, the
shape of theta function for a fixed area argument shows that it is centered
symmetrically around $\eps =0$, and so we will span the
$\eps$-parameter around this value. Also, in order to generate the
mass spectrum one could put the mass matrices in the form \be
M^{u,d}_{ij}  = h_{qu} A v^{u,d}_3 \left(
\begin{array}{ccc}
v^{u,d}_1/v^{u,d}_3 & \a_1 & \a_2 v^{u,d}_2/v^{u,d}_3\\
\a_2 & v^{u,d}_2/v^{u,d}_3 & \a_1v^{u,d}_1/v^{u,d}_3 \\
\a_1v^{u,d}_2/v^{u,d}_3& \a_2v^{u,d}_1/v^{u,d}_3 & 1
\end{array}
\right)  \ee where $\a_1=B/A$ and $\a_2=C/A$. So
one could generate the spectrum provided that \bea \{v_{1}^{u},\,
v_{2}^{u},\, v_{3}^{u}\} & \propto  & \{m_{u},\, m_{c},\, m_{t}\}
\ ,
\nonumber\\
\{v_{1}^{d},\, v_{2}^{d},\, v_{3}^{d}\} & \propto  & \{m_{d},\,
m_{s},\, m_{b}\} \label{proportional} \ .
\eea
and $\a_1,\a_2 \ll 1$. The conditions (\ref{proportional}) with the constraint
(\ref{higgscondition}) would determine the range in which we should
vary the VEVs. With such considerations one finds that the following choice of
parameters:
\bea
 v^u_1  \simeq \ 63\   MeV, &v^u_2 \simeq  \ 0.95\   GeV, & v^u_3 \simeq \ 174\   GeV \nonumber \\
v^d_1  \simeq  \ 8.5\   MeV, &v^d_2 \simeq  \ 136\   MeV, & v^u_3 \simeq \ 4.2\   GeV \nonumber \\
\eps \simeq \ 0.002\ , &area \simeq  \ 18.71\  . \label{masseparameters} \eea
gives the following quarks mass spectrum \bea d_U&=& \{m_t\ =\
173.9\ GeV,\ \ m_c\ =\ 1.02\  GeV,\ \ m_u\ =\ 4.3 \
MeV\}\nonumber\\d_D&=& \{m_b\ =\  4.19\ GeV,\ \ m_s\ =\ 136\ MeV,\
\ m_d\ =\
 8.2\ MeV\}\label{massresult}\eea
which are in the experimentally acceptable range
 \cite{abel3}, and a CKM matrix with diagonal elements near the unity,
and $(V_{CKM})_{12} \simeq 0.216$.
However, $(V_{CKM})_{13} \sim (V_{CKM})_{23} \sim 10^{-4} - 10^{-5}$.

One can also look for other
structures different from the ``hierarchical" $(\a \ll 1)$ texture, for example the
case of  $\a_{1,2} \sim v_1/v_3 \sim v_2/v_3 \sim 1$ leads to a nearly
democratic Yukawa texture which is known to accommodate the observed masses
and mixing. However, our checks with real-valued
vevs indicate that our configuration leads to the correct masses and one
mixing angle only while the other two mixing angles are smaller than
their experimental values. This is similar to $Z_3$-heterotic situation
in \cite{abel}, where another mechanism, Fayet-Iliopoulos breaking, was called
for in order to address the question of the complete quark mixing.
However, with complex vevs and a democtratic texture, one gets the following USY texture.
\be
Y^{u, d}  =  \lambda_{u,d} \left(
\begin{array}{ccc}
e^{i\varphi_{13}^{u,d}} & 1 & e^{i\varphi_{23}^{u,d}}\\
1 & e^{i\varphi_{23}^{u,d}}& e^{i\varphi_{13}^{u,d}} \\
e^{i\varphi_{23}^{u,d}}& e^{i\varphi_{13}^{u,d}} & 1
\end{array}
\right)
\ee
This type of Yukawa, where all Yukawa couplings have the same modulus and the flavor
dependence being all contained in the phases, has been recenetly studied in \cite{branco}.
It was shown that with very small values of the phases
$\sim 10^{-3}-10^{-2}$ one could generate the right values of the
quark masses and the CKM mixing angles. It is interesting to note that this class of couplings
is motivated by horizontal symmetries \cite{branco} and also arises in the models with two large extra
dimension \cite{branco7}. Here we find another motivation for the USY couplings.

\section{{\large \bf Conclusions}}
We have shown how simple configurations of D-branes wrapping a
compact space may give a good quantitatively description of quark
masses and mixing. In particular, one finds that with a 3
supersymmetric Higgs doublets model the anomaly cancellation
condition could be solved easily without introducing extra matter
doublet fields, nor putting assumptions on the quarks' brane
assignment or on the branes themselves. In this class of models,
it turns out that the string scale is of order $10^{12}$ GeV which is
an interesting scale for generating neutrino masses and many other phenomenological
issues. With real Higgs vevs, the model can easily account for the quark
masses and one of the CKM mixing angles. However, with complex vevs one can get
Yukawa couplings in the form of USY textures which can accommodate the masses and the
three CKM mixing angles with very small phases. It would be worthwhile to study the leptonic
sector in this perspective.

\section*{{\large \bf Acknowledgements}}
 The work of S.K. was supported by PPARC. Major part of this work was
done within the Associate Scheme of ICTP. We thank F. Marchesano,
L.E. Ibanez and R. Blumenhagen for useful discussions.


\noindent

\begin{thebibliography}{99}
\bibitem{abel3}K. Hagiwara et al. (Particle Data
Group), {\it Phys. Rev.} {\bf D66} (2002) 010001.

\bibitem{IbTo1} H.~Fritzsch and Z.~z.~Xing,
{\em ``Mass and flavor mixing schemes of quarks and leptons,''}
Prog.\ Part.\ Nucl.\ Phys.\  {\bf 45}, 1 (2000), hep-ph/9912358.
\\
F.~J.~Gilman and Y.~Nir, {\em ``Quark Mixing: The CKM Picture,''}
Ann.\ Rev.\ Nucl.\ Part.\ Sci.\  {\bf 40}, 213 (1990).

\bibitem{IbTo6}
L.~E.~Ib\'a\~nez and G.~G.~Ross,
Phys.\ Lett.\ B {\bf 332}, 100 (1994), hep-ph/9403338.
\\
P.~Bin\'etruy and P.~Ramond,
Phys.\ Lett.\ B {\bf 350}, 49 (1995),
hep-ph/9412385.

\bibitem{IbTo7}
S.~Raby, {\em``Introduction to theories of fermion masses,''}
hep-ph/9501349.

\bibitem{abel7}
L.J. Dixon, J. Harvey, C. Vafa and E. Witten,
{\it Nucl. Phys.} {\bf B261} (1985) 678;
{\it Nucl. Phys.} {\bf B274} (1986) 285.

\bibitem{abel8}
L.E. Ib\'a\~nez, H.P. Nilles and F. Quevedo,
{\it Phys. Lett.} {\bf B187} (1987) 25.

\bibitem{abel9to10}
S. Hamidi and C. Vafa,
{\it Nucl.
Phys.} {\bf B279} (1987) 465;

\noindent
L.J. Dixon, D. Friedan, E. Martinec and S. Shenker,
{\it Nucl. Phys.} {\bf B282} (1987)
13.

\noindent L.E. Ib\'a\~nez,
{\it Phys. Lett.} {\bf
B181} (1986) 269.

\bibitem{abel11to12}
J.A. Casas and C. Mu\~noz,
{\it Nucl. Phys.} {\bf B332} (1990) 189
[Erratum, ibid. {\bf B340} (1990) 280].

\noindent J.A. Casas, F. G\'omez and C. Mu\~noz,
{\it Phys. Lett.}
{\bf B292} (1992) 42.

\bibitem{abel13to14}
J.A. Casas, F. G\'omez and C. Mu\~noz,
{\it Phys. Lett.} {\bf
B251} (1990) 99.

\noindent T.T. Burwick, R.K. Kaiser and H.F. M\"uller,
{\it Nucl. Phys.}
{\bf B355} (1991) 689.

\bibitem{abel15to16}
T. Kobayashi and N. Ohtsubo,
{\it Int. J. Mod. Phys.} {\bf A9} (1994)
87.

\noindent J.A. Casas, F. G\'omez and C. Mu\~noz,
{\it Int. J. Mod. Phys.}
{\bf A8} (1993) 455.

\bibitem{kane6}
{A. Faraggi, D.V. Nanopoulos, and K. Yuan, \NPB{335}{1990}{347};
A. Faraggi, \PRD{46}{1992}{3204};A. Faraggi, \PLB{278}{1992}{131},
\NPB{403}{1993}{101} and \PLB{339}{1994}{223};G. Cleaver, A.
Faraggi, D.V. Nanopoulos, and J. Walker, hep-ph/9910230.}

\bibitem{kane4and8}
{L.~Ib\'a\~nez, J.E.~Kim, H.P.~Nilles and F.~Quevedo,
\PLB{191}{1987}{282}; J.A.~Casas and C.~Mu\~noz,
\PLB{209}{1988}{214} and \B{214}{1988}{157}; J.A.~Casas,
E.~Katehou and C.~Mu\~noz, \NPB{317}{1989}{171}; A.~Font,
L.~Ib\'a\~nez, H.P.~Nilles and F.~Quevedo, \PLB{210}{1988}{101};
A.~Chamseddine and M.~Quir\'os, \PLB{212}{1988}{343},
\NPB{316}{1989}{101}; A.~Font, L.~Ib\'a\~nez, F.~Quevedo and
A.~Sierra, \NPB{331}{1990}{421}.}

\noindent {N. Irges, S. Lavignac, P. Ramond,
\PRD{58}{1998}{035003};P. Binetruy, S. Lavignac, P. Ramond,
\NPB{477}{1996}{353}.}

\bibitem{kane12}
{G. Pradisi and A. Sagnotti, \PLB{216}{1989}{59}; M. Bianchi and
A. Sagnotti, \PLB{247}{1990}{517}, \NPB{361}{1991}{519}; E. Gimon
and J. Polchinski, \NPB{477}{1996}{715}, hep-th/9601038; C.
Angelantonj, M. Bianchi, G. Pradisi, A. Sagnotti and Ya.S. Stanev,
\PLB{385}{1996}{96}, hep-th/9606169; C. Angelantonj, M. Bianchi,
G. Pradisi, A. Sagnotti, and Ya.S. Stanev, \PLB{387}{1996}{743}.}


\bibitem{kane14}
{Z. Kakushadze, G. Shiu and S.-H. Tye, \PRD{58}{1998}{086001},
hep-th/9803141; M. Berkooz and R.G. Leigh, \NPB{483}{1997}{187},
hep-th/9605049; G. Zwart, \NPB{526}{1998}{378}, hep-th/9708040; Z.
Kakushadze, \NPB{512}{1998}{221}, hep-th/9704059; Z. Kakushadze
and G. Shiu, \PRD{56}{1997}{3686}, hep-th/9705163;
\NPB{520}{1998}{75}, hep-th/9706051; L.E. Ib\'a\~nez,
hep-th/9802103; D. O'Driscoll, hep-th/9801114; J. Lykken, E.
Poppitz, and S. Trivedi, \NPB{543}{1999}{105}, hep-th/9806080.}

\bibitem{kane15to17}
{G. Shiu and S.-H. Tye, \PRD{58}{1998}{106007}, hep-th/9805157.}

\noindent {G. Aldazabal, L. Ib\'a\~nez, and F. Quevedo,
hep-ph/0001083; JHEP 0001:031,2000, hep-th/9909172; G. Aldazabal,
L. Ib\'a\~nez, F. Quevedo, and A. Uranga, hep-th/0005067.}


\noindent {M. Cveti\v{c}, M. Plumacher, and J. Wang,
hep-th/9911021.}

\bibitem{khalil33}
{L. Ib\'a\~nez, C. Mu\~noz, and S. Rigolin, \NPB{553}{1999}{43},
hep-ph/9812397.}

\bibitem{khalil35}
M. Berkooz and R.G. Leigh, Nucl. Phys. B483 (1997) 187.

\bibitem{kane19}
{M. Brhlik, L. Everett, G. L. Kane, and J. Lykken, hep-ph/9905215,
hep-ph/9908326.}

\bibitem{kane20}
{E. Accomando, R. Arnowitt, B. Dutta,\PRD{61}{2000}{075010}; S.
Khalil [hep-ph/9910408]; T.Ibrahim and P. Nath,
\PRD{61}{2000}{093004}.}

\bibitem{khalil}
D.G. Cerdeno, E. Gabrielli, S. Khalil, C. Munoz, E.
Torrente-Lujan, {\it ``Determination of the string scale in
D-brane scenarios and dark matter implications"},
\NPB{603}{2001}{231}

\bibitem{kane}
L. Everett, G.L. Kane and S.F. King,{\it ``D Branes and
Textures"}, JHEP 0008 (2000) 012.


\bibitem{IbMo1and4}
R.~Blumenhagen, L.~G\"orlich, B.~K\"ors and D.~L\"ust, \
Fortsch.\ Phys.\  {\bf 49}, 591 (2001), hep-th/0010198.

\noindent R.~Blumenhagen, B.~K\"ors and D.~L\"ust, JHEP {\bf
0102}, 030 (2001), hep-th/0012156.

\bibitem{aldazabal1}
G.~Aldaz\'abal, S.~Franco, L.~E.~Ib\'a\~nez, R.~Rabad\'an and
A.~M.~Uranga, J.\ Math.\ Phys.\  {\bf 42}, 3103 (2001),
hep-th/0011073.


\bibitem{IbGe}
L.~E.~Ib\'a\~nez, F.~Marchesano and R.~Rabad\'an, {\em ``Getting
just the Standard Model at Intersecting Branes,''} JHEP {\bf
0111}, 002 (2001), hep-th/0105155.

\bibitem{kokorelis}
C. Kokorelis, JHEP {\bf 0209},(2002) 029, hep-th/0205147; {\bf
0208},(2002) 036, hep-th/0206108; hep-th/0309070

\bibitem{cevetic}
M.~Cveti\v c, G.~Shiu and A.~M.~Uranga,
Phys.\ Rev.\ Lett.\  {\bf 87}, 201801 (2001), hep-th/0107143.
Nucl.\ Phys.\ B {\bf
615}, 3 (2001), hep-th/0107166.

\bibitem{IbYu}
D. Cremades, L. E. Ibanez and F. Merchesano,{\it ``Yukawa
couplings in intersecting D-brane models"}, JHEP 0307 (2003) 038.

\bibitem{IbTo}
D. Cremades, L. E. Ibanez and F. Merchesano,{\it ``Towards a
theory of quark masses, mixings and CP-violation"}, hep-ph/0212064

\bibitem{Alday}
L. F. Alday and G. Aldazabal, JHEP {\bf 0205}, 022 (2002),
hep-th/0105155.

\bibitem{abel20}
L.E. Ib\'a\~nez, J.E. Kim, H.P. Nilles and F. Quevedo,
{\it Phys. Lett.} {\bf B191} (1987) 3.

\bibitem{abel}
S. Abel and C. Munoz, {\em ``Quark and lepton masses and mixing
angles from superstring constructions''}, JHEP {\bf 02}, 010
(2003),

\bibitem{antonA}
I. Antoniadis, E. Kiritsis and T.N. Tomaras, Phys. Lett. B486
(2000) 186.

\bibitem{antonB}
I. Antoniadis, E. Kiritsis, J. Rizos and T.N. Tomaras, Nucl. Phys.
B660 (2003){81}

\bibitem{anton97}
I . Antoniadis and M. Quiros, Phys. Lett. B392 (1997) 61

\bibitem{khalil11}
K. Benakli, Phys. Rev. D60 (1999) 104002.

\bibitem{khalil12}
C. Burgess, L.E. Iba\~nez and F. Quevedo, Phys. Lett. B447 (1999)
257.

\bibitem{khalil15}
J.A. Casas, A. Lleyda and C. Mu\~noz, Phys. Lett. B389 (1996) 305.

\bibitem{khalil16}
S.A. Abel, B.C. Allanach, F. Quevedo, L.E. Ib\'a\~nez and M.
Klein, JHEP 0012 (2000) 026.

\bibitem{khalil17}
N. Kaloper and A. Linde, Phys. Rev. D59 (1999) 101303.

\bibitem{munoz}
C. Munoz, {\em ``A kind of prediction from superstring model
building''} JHEP {\bf 12}, 015 (2001)

\bibitem{munoz39}
D.E. Groom et al. (Particle Data Group), {\it Eur. Phys. J.} {\bf
C15} (2000) 1.


\bibitem{blumenhagen}
R.~Blumenhagen, L.~G\"orlich, B.~K\"ors and D.~L\"ust, JHEP {\bf
0010}, 006 (2000), hep-th/0007024;

\bibitem{aldazabal2}
\noindent G.~Aldaz\'abal, S.~Franco, L.~E.~Ib\'a\~nez,
R.~Rabad\'an and A.~M.~Uranga, JHEP {\bf 0102}, 047 (2001),
hep-ph/0011132.

\bibitem{IbYu27}
A.~M.~Uranga,
Nucl.\ Phys.\ B {\bf 598}, 225 (2001),
hep-th/0011048.

\bibitem{IbYu12}
D.~Cremades, L.~E.~Ib\'a\~nez and F.~Marchesano,
JHEP {\bf 0207}, 022 (2002), hep-th/0203160.

\bibitem{IbYu17}
A.~M.~Uranga,
JHEP {\bf 0212}, 058 (2002), hep-th/0208014.

\bibitem{branco}
G.C. Branco, M.E. Gomez, S. Khalil and A.M. Teixeira, Nucl.\
Phys.\ B {\bf 659}, 119 (2003).

\bibitem{branco7}
P.~Q.~Hung and M.~Seco,
Nucl.\ Phys.\ B {\bf 653}, 123 (2003)
[arXiv:hep-ph/0111013].





\end{thebibliography}
\end{document}